\documentclass[
 reprint,
showpacs,
 amsmath,amssymb,
 aps,
pra,
]{revtex4-1}

\usepackage{graphicx}
\usepackage{dcolumn}
\usepackage{bm}
\usepackage{hyperref}

\def\v#1{{\bf#1}}
\def\be{\begin{equation}}
\def\ee{\end{equation}}
\def\bea{\begin{eqnarray}}
\def\eea{\end{eqnarray}}

\def\ncal{\mbox{$\cal N\,$}}
\def\fcal{\mbox{$\cal F\,$}}
\def\<{\langle}
\def\>{\rangle}

\begin{document}


\title{Diffraction of particles in free fall}
\author{D. Condado$^1$}
\author{J. L. D\'iaz-Cruz$^1$}
\author{A. Rosado$^2$}
\author{E. Sadurn\'i$^2$}
 \email{sadurni@ifuap.buap.mx}
\affiliation{$^1$Facultad de Ciencias F\'isico Matem\'aticas, Benem\'erita Universidad Aut\'onoma de Puebla, 72570 Puebla, M\'exico}
\affiliation{$^2$Instituto de F\'isica, Benem\'erita Universidad Aut\'onoma de Puebla,
Apartado Postal J-48, 72570 Puebla, M\'exico}

\date{\today}

\begin{abstract}
The problem of a beam of quantum particles falling through a diffractive screen is studied. The solutions for single and double slits are obtained explicitly when the potential is approximated by a linear function. It is found that the resulting patterns depend on a quasi-time $\tau$ given by a function of the coordinate along the propagation axis in a classical combination $z_0-t^2 F/m$, while the diffraction effects along transverse axes are due solely to $m/\hbar$. The consequences on the precision at which the equivalence principle can be tested are discussed. Realizations with ultra cold neutrons, Bose-Einstein condensates and molecular beams are proposed.
\end{abstract}



\maketitle


\section{ Introduction \label{sec:1}}

Over the last forty years \cite{colella}, free falling quantum mechanical particles of different kinds \cite{nesvi1,nesvi2,nesvi3,bateman,aguilera,muntinga,vanzoest} have been subjected to various types of perturbations with the purpose of testing the Principle of Equivalence \cite{lammerzahl, synge, wald}. Due to the many physical limits and layers in which Einstein's principle can be tested, it is convenient here to quote a standard text (\cite{weinberg1972}, pp. 68-69) containing at least three important distinctions: 1. Galileo's Universality of Free Fall (UFF), 2. The Weak Equivalence Principle (WEP) and 3. The Strong Equivalence Principle (SEP). Our aim in this paper is to discuss the role of diffraction in such situations. 

For convenience and clarity, let us state these principles in the following forms:

\begin{itemize}
\item[1.] Under the influence of a gravitational field, different test masses experience the same accelerations and, under the same initial conditions, they undergo equal trajectories (equivalent to UFF).
\item[2.] The gravitational and the inertial mass of a body are equal (equivalent to WEP). See our comment \cite{proportionality} on proportionality between masses.
\item[3.] For all local inertial frames, the laws of physics are equivalent to those of unaccelerated systems without gravitation (SEP). This extends beyond freely falling bodies, e.g. electromagnetic radiation. 
\end{itemize}

UFF and WEP are seemingly equivalent in classical mechanics, where it is possible to show that 2 implies 1 due to a cancellation of masses $m_i=m_g$ in the equations of motion. Principle 2 is further generalized to relativity, when trajectories of different test masses are postulated as geodesics in a background space-time (and such is the form of 3). Any classical deviation from 1 using point masses necessarily entails a violation of 2, but such a historically recognized equality \cite{galileo} has been experimentally tested with great accuracy since the times of E\"otv\"os \cite{eotvos}.

It is now well-known \cite{colella} that this situation is no longer true when we deal with quantum-mechanical wave packets, even in a non-relativistic regime. The explanation in such a scenario is quite simple: Although the Heisenberg equations of motion may not violate 2, different inertial masses $M_i, m_i$ may produce different diffusivities of wave packets --e.g. Gaussian distributions \cite{gaussianevolution}-- appearing in combinations $\hbar/m_i$ or $\hbar/M_i$ in front of our natural time units. Thus, even in the absence of a gravitational field, different masses lead to unequal undulatory effects.

In this work we treat the problem of a falling particle passing through a diffractive screen of a general nature and we specialize it to single and double slit configurations, see figure \ref{fig:0}. The method consists in obtaining a stationary diffraction pattern (i.e. a beam with well defined energy $E$) in the presence of a potential along the axis of propagation ($z$ axis). This shall relate the corresponding $z$ coordinate along the fall with a quasi-time and, as expected, it will produce a modified diffraction pattern when compared to those emerging without external fields -- or patterns obtained at rest, should we choose a coordinate frame moving with the average position. It will be shown that the quasi-time depends on both the field intensity $F/m_i$ and the diffusivity $\hbar/m_i$. In this setting, we shall find that:

\begin{figure*}[t]
\includegraphics[height=9cm]{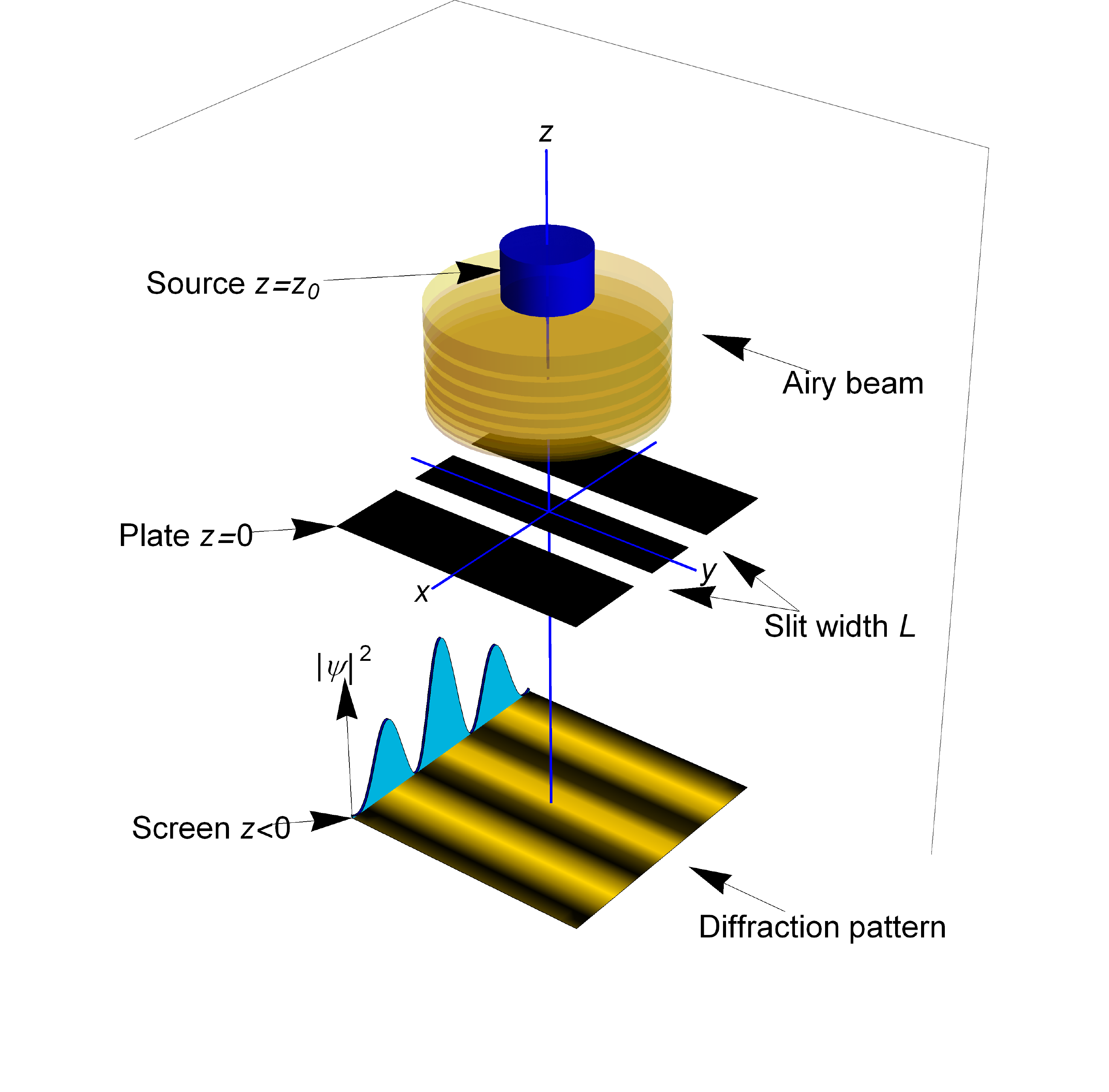}
\includegraphics[width=8cm]{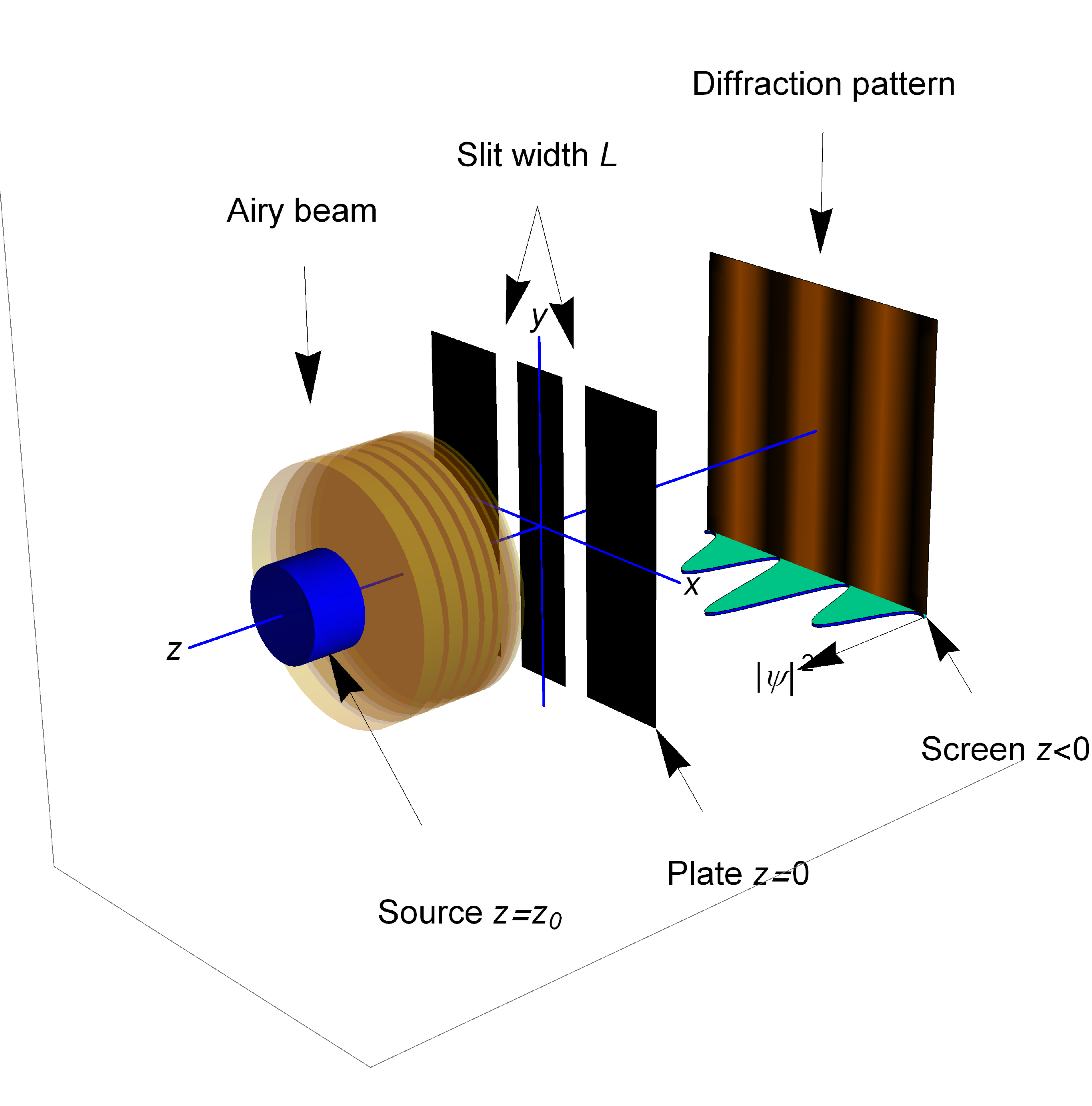}
\caption{\label{fig:0} A diagram of our physical system. The nature of beam sources at $z=z_0$ (blue) is discussed in section \ref{sec:5}. The stationary incident wave (yellow) is given by Airy functions, due to gravity. The diffraction plate at $z=0$ (black) is the quantum analogue of opaque screens in optics, e.g. Cadmium plates for neutrons. Detectors of adequate resolution are placed at $z<0$.}
\end{figure*}

\begin{itemize}
\item[1'.] Diffraction patterns and phase factors are different along the fall if $m_i \neq M_i$. This can be used as a test of 1 if we compare waves below the screen that emerge from identical square packets produced by slits. The energies of monochromatic beams must be such that $E'= M_i E/m_i $ for a fair comparison.

\item[2'.] Diffraction patterns are modified by deviations of $m_i = m_g$. This can be used as a test of principle 2. In parallel, a variation of $g$ in surface gravity will produce similar modifications, which can be used for gravitometry.


\end{itemize}

Regarding the quantum-mechanical grounds of our study, diffraction problems of matter waves have been extensively studied in the literature \cite{moshinsky52, manko1999}. Two well-known reports account for this body of work \cite{kleber1994, delcampo2009}. The reader may benefit from these references to put the work in
context. 

Structure of this paper: In section \ref{sec:2} we recall the specific dependence of wave dynamics on mass ratios and Planck's constant in known examples. Specifically, subsection \ref{sec:2.1} deals with quantum bounces, subsection \ref{sec:2.2} with propagators and \ref{sec:2.3} deals with phase shifts. Our original treatment of a free falling beam of particles starts in section \ref{sec:3}, where the diffraction problem in a force field is solved semi classically under the assumption of paraxiality. In section \ref{sec:4}, the formal solution of the problem without approximations is presented using an expansion in Airy functions and a wave falling near the diffraction plate is studied. In section \ref{sec:5} we discuss beam realizations in nuclear, molecular and atomic physics with an emphasis on the locus of a focal point in diffraction patterns. We conclude briefly in \ref{sec:6} and the numerical estimate of the focal point is left to the Appendix \ref{sec:a1}.

\section{ Mass ratios in trajectories, propagators and interferometers \label{sec:2}}

The aim of this section is to exhibit the effects of gravity in free-falling wave packets, treated in previous works. We divide the discussion of this section in three types of physical situations: Quantum bounces, wave packet propagation without boundaries, and interference effects with phase shifts. In general, we have the following setting: The Schr\"odinger equation of a particle under the effects of a static gravitational source is

\bea
\left[-\frac{\hbar^2}{2 m_i} \nabla^2 + V_{\rm Grav}(\v x)\right] \psi(\v x; t)= i \hbar \frac{\partial \psi(\v x ; t)}{\partial t}.
\label{0.1}
\eea
For experiments carried out in limited vertical distances, it is reasonable to approximate $V_{\rm Grav}(\v x) \approx Fz$, where $F = m_g \times g$, $m_g$ is the gravitational mass and $g$ the gravitational acceleration at the particular height of our experiment. It should be noted that a strong similarity with a constant electric field may arise with this approximation; however, the electromagnetic emulation of a gravitational field is typically avoided due to the lack of an equivalence principle for different charges. Equal charge-to-mass ratios for all particles are not sufficiently general conditions. When (\ref{0.1}) is viewed as an initial value problem, the wave describing a falling particle satisfies

\bea
\psi(\v x; t) = \int d^3 x' K(\v x,\v x'; t) \psi(\v x'; 0),
\label{0.2}
\eea
where the propagator $K$ has a Gaussian form. This is mathematically equivalent to any problem with linear (or even quadratic) potentials and $K$ can be found in many different ways \cite{akhundova85, farina86, feynman2005}. In our problem, the factors $F/m_i$ and $m_i/\hbar$ shall appear explicitly.

\subsection{Relevant ratios in quantum bounces within the WKB approximation \label{sec:2.1}}

When the contact with the ground after the fall is taken into account, eq. (\ref{0.1}) must be modified by the triangular potential 

\bea
V(z) = \begin{cases} Fz & \text{if}\quad z>0 \\ \infty & \text{if}\quad z<0. \end{cases}
\label{0.3}
\eea
This gives rise to bound states within the potential well. We may resort to WKB approximations to give an idea of how the relevant factors $F/m_i$, $\hbar/m_i$ appear in energy differences.
Several experiments have been carried out to show the effects in question. Quantum bounces of free falling neutrons can have access to the violation of both principles 1 and 2 \cite{nesvi1, nesvi2, nesvi3}. In a simple semi classical calculation --also possible beyond semiclassics using the roots of the Airy function-- the Bohr frequency between two stationary states is
\bea
&&\frac{E_n-E_{n'}}{\hbar} = \left( \frac{m_i}{\hbar} \right)^{1/3} \left( \frac{F}{m_i} \right)^{2/3} \times \nonumber \\ &\times &   \left\{ \left[\frac{3\pi}{8}(n-1/4) \right]^{2/3}-\left[ \frac{3\pi}{8}(n'-1/4)\right]^{2/3} \right\}
\label{1}
\eea 
so that any perturbation capable of driving such transitions would produce a period of oscillation containing $F/m_i = g m_g / m_i$ and $\hbar / m_i$. The quantum dynamics of a more general wave entails the appearance of all Bohr frequencies of the form (\ref{1}), showing the same dependence for all Fourier components:

\bea
|\psi(\v x;t)|^2 = \sum_{n,n'} C_n C_{n'}^* \psi_n(\v x) \psi_{n'}^*(\v x) e^{-i\omega_{n,n'}t}
\label{fourier}
\eea
where $\omega_{n,n'} \equiv (E_n-E_{n'})/\hbar$ and $C_n = \< \psi_n | \psi,0\>$.

\subsection{Relevant ratios in propagators \label{sec:2.2}}

The relevant ratios that contain possible deviations from $m_i=m_g$ in (\ref{1}) also appear in the problem without bounces. We turn now to the initial value problem in (\ref{0.2}) with known Gaussian propagators. In this setting, $K$ can be factorized in three parts $K=K_x K_y K_z$. We shall see that if an obstacle, such as a diffraction plate, is placed in the fall, this factorization is no longer valid.
For an arbitrary wavepacket falling from the leaning tower of Pisa, the propagator in the $z$ coordinate and without bounces fulfills the relation \cite{grosche}
\bea
K_{{\rm field\ }} (z,z';t)&=& K_{{\rm free\ } }(z-F t^2 /2m_i,z';t)\times \nonumber \\ &\times & \exp\left\{ -i\frac{m_i }{\hbar} \left[ \left(\frac{F}{m_i}\right) zt - \left(\frac{F}{m_i} \right)^2 \frac{t^3}{6} \right] \right\}\nonumber \\
\label{2}
\eea
%
which gives rise to wave dynamics of the type
\begin{widetext}
\bea
\psi_{{\rm field\ } } (x,y,z;t)= \exp\left\{ -i \left(\frac{m_i }{\hbar} t \right) \left(\frac{F}{m_i}\right)  \left[  z - \left(\frac{F}{m_i} \right) \frac{t^2}{6} \right] \right\} \psi_{{\rm free\ } }(x,y,z-Ft^2/2m_i;t).
\label{3}
\eea
\end{widetext}
This means that in a very general picture, the analysis of a single packet boils down to free undulatory effects plus a translation along the path $z(t)= z- t^2 F/2m_i$ for the density and a translation along $z(t)= z- t^2 F/6m_i$ for the phase. This is in compliance with (\ref{1}) and shows that principle 2 breaks down only in a classical fashion, while the factor $\hbar/m_i$ is the sole responsible for quantum effects and the breakdown of principle 1. 

\subsection{Relevant ratios in interferometry and phase shifts \label{sec:2.3}}

The phase factor in (\ref{3}), on the other hand, has been exploited in Mach-Zehnder interferometers, where different paths with equal end points would yield the following phase difference 

\bea
\Delta \Phi = \left(\frac{m_i}{\hbar} \right)^2 \left(\frac{F}{m_i} \right) \times A,
\label{4}
\eea
with $A$ the area enclosed by the two paths \cite{chiow}. One can use the Feynman formulation of the propagator to make the classical action appear in (\ref{4}), so this calculation is quite general. Once more, the sensitivity in the non-relativistic regime of these experiments is bound to the factors $F/m_i$ and $\hbar/m_i$. Factors depending on proper times along two paths may be achieved if relativistic corrections were included; this can be consulted in \cite{greenberger}.

In all, the solutions that we are seeking for in a free falling diffraction setting should exhibit similar field and mass dependences as those in (\ref{1}, \ref{2}) and (\ref{3}). The surplus is that the resulting pattern shall be static and easier to measure in an experiment, since the only requirement for detection is a good fluorescent screen at various heights $z$.

\section{ Stationary diffraction patterns in an external field \label{sec:3}}

In some diffraction problems, space and time variables are analogous, thus they can be solved in the same way. Examples of stationary diffraction patterns treated as effective time-dependent problems with square packets can be found in \cite{godoy2002, delcampo2006, sadurni2012}. However, in the presence of potentials that depend on the coordinate along the propagation axis, such equivalence is no longer valid. For this reason, we present our treatment of diffraction in space from scratch. 
 
A neutron falling from a source into a diffractive plate at $z=0$ is subjected to a force given by $-V'(z)$. In the vicinity of the screen, the particle cannot be represented by a plane wave, but instead by the stationary solution of 

\bea
\left[- \frac{\hbar^2}{2m_i}\left( \partial_x^2 + \partial_z^2 \right) + V(z) \right] \psi(x,z;E)=E\psi(x,z;E), \nonumber \\
\label{1.1}
\eea
where $y$ is ignorable if the plate has slits. In a semi classical approach, if the gravitational potential $V$ does not have strong variations, we may choose for $z>0$ the WKB wave

\bea
\psi(x,z) &=& \ncal \frac{\exp\left(-i\int^z_{0} k(\eta) d\eta \right)}{\sqrt{|k(z)|}}, \nonumber \\  k(z) &=& \sqrt{\left( \frac{2m_i}{\hbar^2} \right)(E-V(z))}.
\label{1.2}
\eea
The singular density at the turning point can be avoided by a better approximation, but in what follows this will be irrelevant, as the WKB phase shall play the important role. Below the screen we have a solution of (\ref{1.1}) that depends on both $x$ and $z$. First we absorb the propagating factor with the transformation

\bea
\tilde \psi(x,z) = \psi(x,z)\exp\left(+i\int^z_{0} k(\eta) d\eta \right).
\label{1.3}
\eea
This turns the Schr\"odinger equation into
\bea
&&\left[- \frac{\hbar^2}{2m_i}\partial_x^2 + i \frac{\hbar^2 k(z)}{m_i} \partial_z \right] \tilde \psi(x,z;E)= \nonumber \\ &&\frac{\hbar^2}{2m_i} \left[\partial_z^2 -ik'(z) \right] \tilde \psi(x,z;E),
\label{1.4}
\eea
and $k'(z)= im_i V'(z)/\hbar^2 k(z)$. The semiclassical approach in $z$ assumes a large $k(z)$, a small $V'(z)$ and small second derivatives of $\tilde \psi$, which can be established as

\bea
\left|\tilde \psi_{zz} + \frac{im_i V'(z)}{ \hbar^2 k(z)}\tilde \psi \right| \ll \left|k(z)\tilde \psi_z \right|.
\label{1.5}
\eea
From here the r.h.s. of (\ref{1.4}) can be safely neglected. Then we apply the quasi time transformation \cite{footnote}

\bea
\tau(z) = - \int_{0}^{z} \frac{m}{\hbar k(\eta)} d\eta, \quad z(\tau) = - \int_{0}^{\tau} \frac{\hbar k}{m} dt,   
\label{1.6}
\eea
which is more than a cosmetic change, since it casts the stationary diffraction problem (\ref{1.1}) into an effective one-dimensional time-dependent propagation problem; a clean expression is obtained:

\bea
\left[- \frac{\hbar^2}{2m_i}\partial_x^2 - i \hbar \partial_{\tau} \right]\tilde \psi(x,z(\tau)) = 0.
\label{1.7}
\eea
In the classically allowed region the quasi time is, in principle, a real quantity; we shall see that the effect of tunneling can be described as well if it becomes complex. The wave shall be obtained by analytic continuation of the solutions into the classically forbidden region.

\subsection{ One and two slits in a linear potential \label{sec:3.1} }

The case $V(z)= F z$ with $F>0$ and $\tau$ real yields

\bea
k(z) &=& \sqrt{\left( \frac{2m_i}{\hbar^2}\right) |E-F z|}, \nonumber \\ \tau(z) &=&  \frac{\sqrt{2m_i}}{F} \left( \sqrt{|E-F z|}-\sqrt{|E|} \right).
\label{1.8}
\eea
The initial condition is equivalent to the wave coming out of the screen conformed by plates that absorb impinging particles except at some intervals:

\bea
\psi(x,0)= \tilde \psi(x,z(0)) = \frac{1}{\sqrt{L}} \Theta(L/2 - |x|)
\label{1.9}
\eea
for a slit centred at the origin and

\bea
\psi(x,0)&=& \tilde \psi(x,z(0))\nonumber \\ &=& \frac{1}{\sqrt{2L}} \left[\Theta(L/2 - |x+a|)+\Theta(L/2 - |x-a|) \right]\nonumber \\
\label{1.10}
\eea
for two slits centred at $x=\pm a$ and $a>L/2$. When $a \rightarrow 0$ one reduces to the other. The general case yields the following solution

\bea
\tilde \psi(x,z(\tau))&=& \frac{1}{\sqrt{2L \pi}} e^{-i\pi/4} \left[ \fcal(x_{+,+})+ \fcal(x_{-,-}) \right. \\ \nonumber  &+& \left. \fcal(x_{+,-}) + \fcal(x_{-,+}) \right], \nonumber \\ x_{p,q} &=& \sqrt{\frac{m_i}{2\hbar |\tau(z)|}} (L/2 +p a + q x) 
\label{1.11}
\eea
with $\fcal$ the exponential Fresnel integral

\bea
\fcal(Z) = \int_{0}^{Z} e^{i x^2} dx.
\label{1.12}
\eea
Evidently $|\tilde \psi|=|\psi|$ in the classically allowed region and the overall semiclassical phase factor adds nothing to the intensity pattern, but the influence of $F$ makes itself present in the position of maxima and minima in the new quasi time. We also note that (\ref{1.11}) is not factorizable in $x,z$.

For a single slit, a focusing point along the $z$ axis emerges \cite{case}. The focusing time is such that $ \tau \approx 0.055 \times m_i L^2 / \hbar$, where the approximate numerical constant is in fact a root of a transcendental equation involving Fresnel functions. The details can be found in the Appendix \ref{sec:a1}. Inverting $z$ in terms of $\tau$ using (\ref{1.8}) gives an approximation of the focusing height:
\begin{widetext}
\bea
z_{\rm focus} (E,L, F/m_i,\hbar/m_i) \approx \frac{E}{F} - \frac{Fm_i L^4}{2 \hbar^2} \times \left( 0.055 + \frac{\hbar}{L^2 F} \sqrt{\frac{2|E|}{m_i}} \right)^2.
\label{1.13}
\eea
\end{widetext}
This function seems to depend on $m_i$ and $F$ in extraneous ways, but some physical considerations are in order. When the energy of the emerging beam is controlled by the particle's wavelength or wave number, expression (\ref{1.13}) is indeed useful and exhibits new combinations of mass and force. Nevertheless if a particle abandons a radiant source at a certain temperature (e.g. a furnace or a reactor) we must estimate its initial velocity $v$ from a statistical rapidity distribution and then select a specific velocity experimentally via a rotating shutter or chopper; this forces to consider $E= m_i v^2 /2 + F z_0$ in our non-relativistic experiment. Then (\ref{1.13}) is reduced to a more palatable expression in terms of $\alpha \equiv m_i/F$, $\beta \equiv m_i / \hbar$:
\begin{widetext}
\bea
z_{\rm focus} (v,L, \alpha,\beta) = z_0 + \frac{v^2 \alpha}{2}  - \frac{L^4 \beta^2}{2 \alpha} \left[ 0.055 + \frac{  \alpha \sqrt{v^2 + 2 z_0/\alpha}}{ \beta L^2}  \right]^2.
\label{1.14}
\eea
\end{widetext}
We should note that the factor in square brackets will depend on $m_i$ only if the initial height of the drop is $z_0 \neq 0$, because the ratio $\alpha/\beta$ has no inertial mass. We also identify the first two terms in (\ref{1.14}) as the classical contribution to the focusing height, while the third term is completely quantum-mechanical. As to the appearance of Planck's constant, only the term affected by the numerical value 0.055 contains it. Special cases can be studied using this formula; for instance, if the particle is shot upwards at zero total energy, i.e. $z_0 < 0$, only the quantum term survives. Also, if the energy is properly adjusted, the quantum term can eliminate the classical terms, which can be traced back to the coincidence between the location of the maximum and the classical turning point.

\begin{figure*}[t]
\includegraphics[width=14cm]{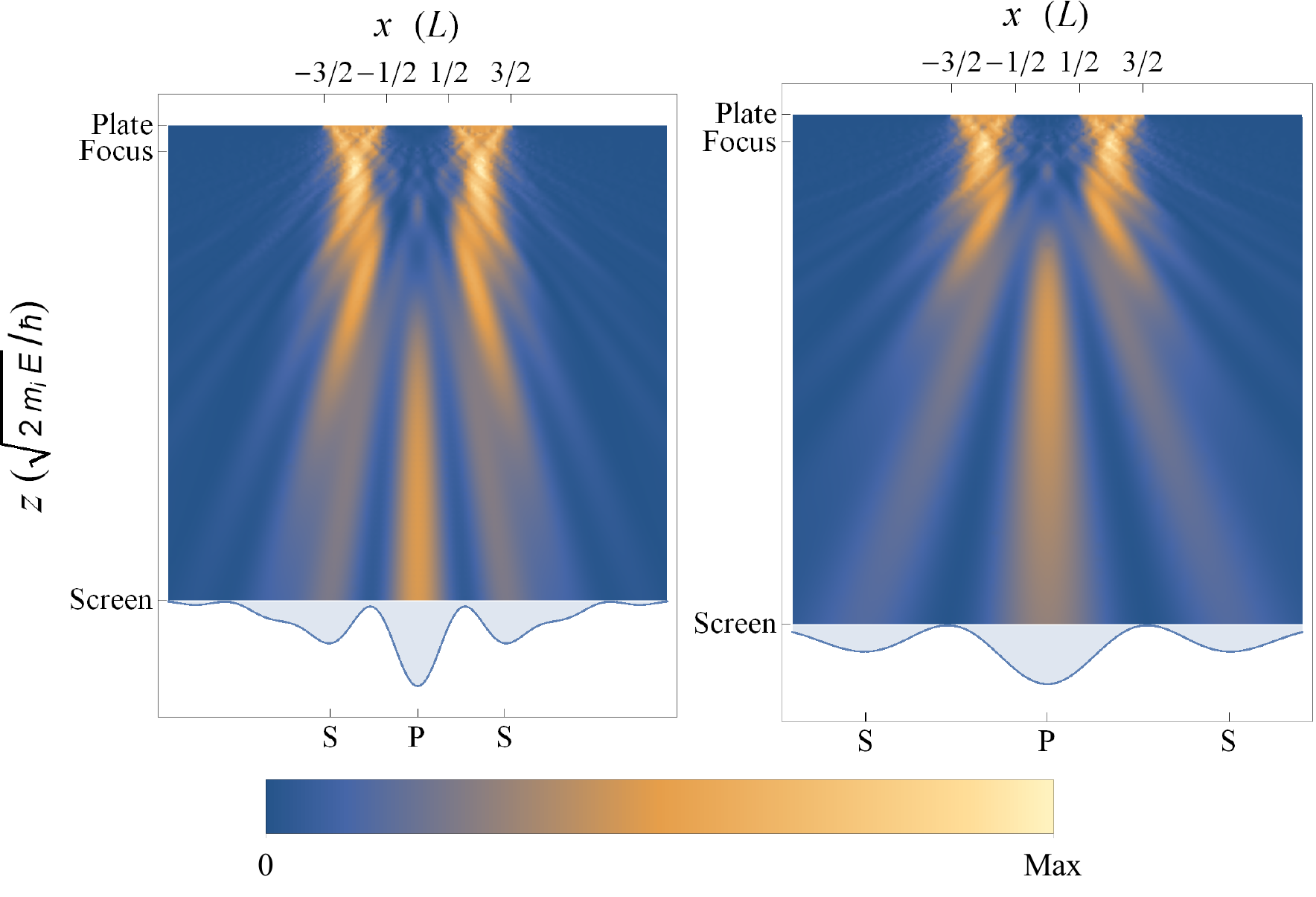} 
\caption{\label{fig:1} Left panel: Double slit diffraction pattern with gravitational pull $F=5, E=2, m_i= m_g = 1$. Right panel: Similar pattern with $F=0$ and same energy $E=2$. The results show an elongation of the shape due to $F$; the mark of the free focus point is kept in both panels as point of reference and the detection screen represents the ground. The distance between primary (P) and secondary (S) maxima at the screen is also modified by $F$. We employ dimensionless variables $z \hbar/ \sqrt{2m_i E}$ (vertical axis) and $x/L$ (horizontal axis).}
\end{figure*}

\begin{figure*}[t]
\includegraphics[width=12cm]{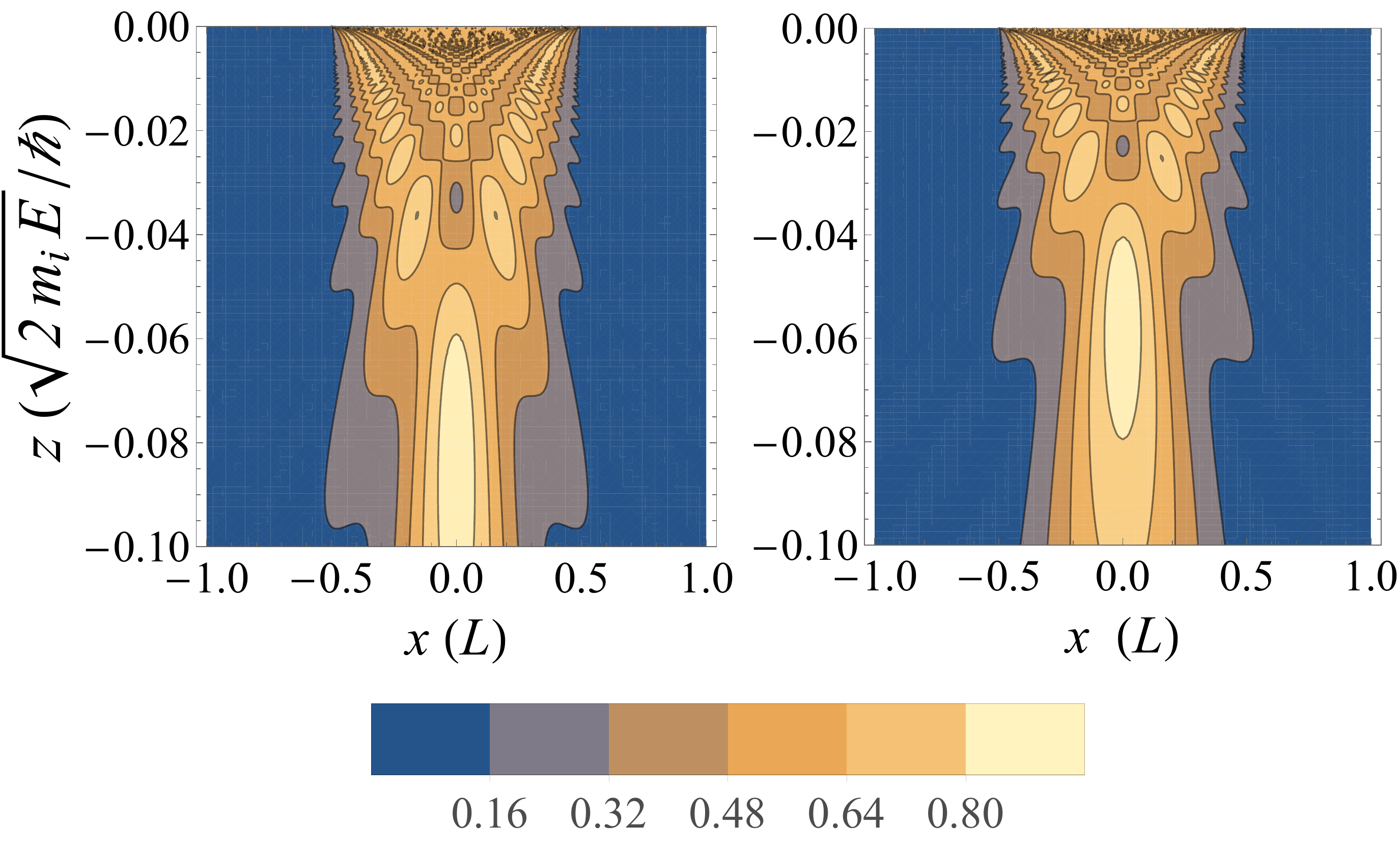} 
\caption{\label{fig:2} Left panel: Single slit diffraction pattern in the near zone, with gravitational pull $F=5, E=2, m_i= m_g = 1$. The focusing point is $z\approx -0.09$ in dimensionless variable. Right panel: Same pattern without gravitational pull and same energy $E=2$. Once more we see an elongation with respect to the focus at $-0.055$. However, the intricacies due to plate's edges are comparable, showing that interference effects are of the same nature with or without $F$. We employ dimensionless variables $z \hbar/ \sqrt{2m_i E}$ (vertical axis) and $x/L$ (horizontal axis).}
\end{figure*}

\begin{figure*}[t]
\includegraphics[width=14cm]{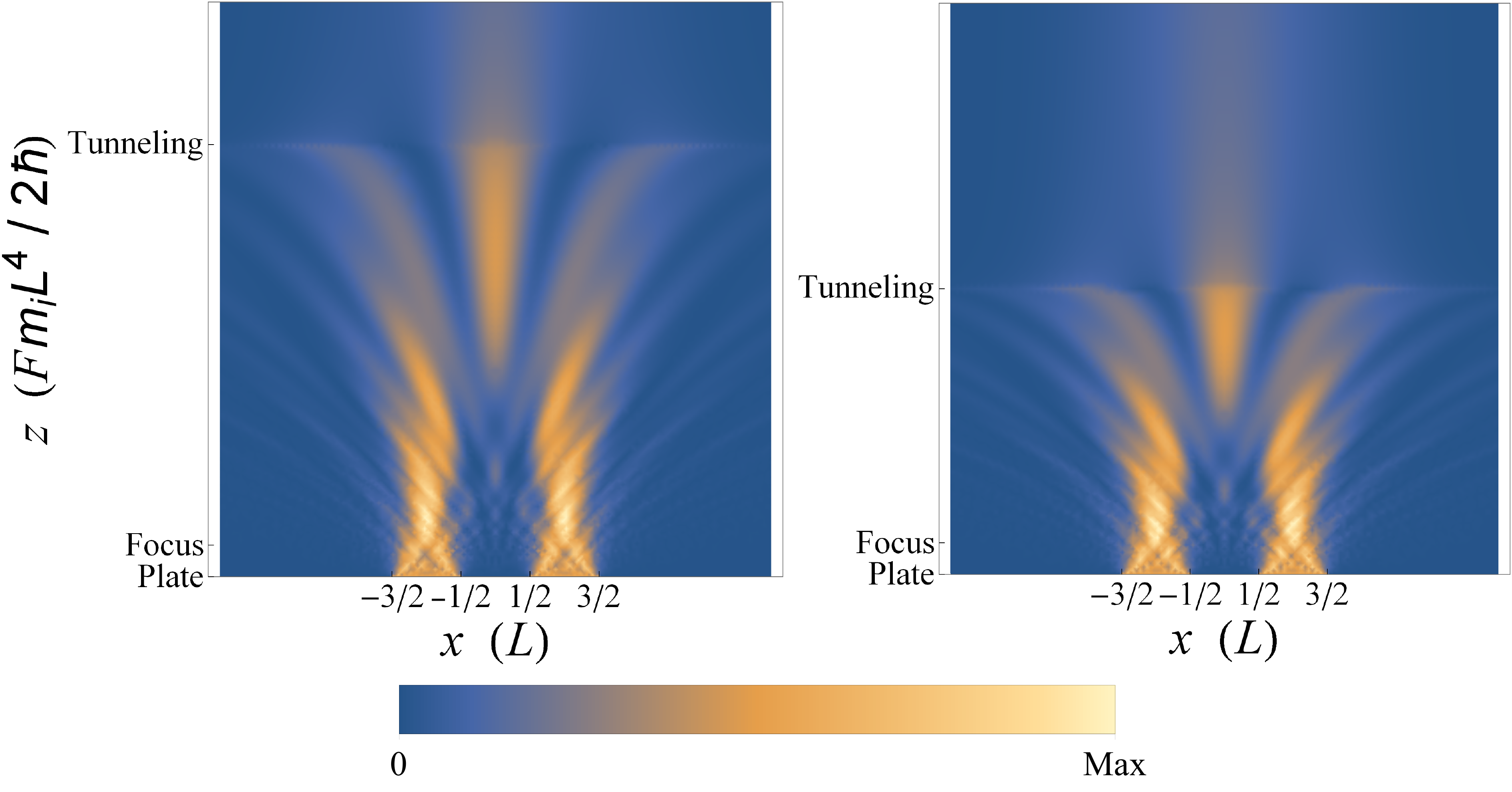} 
\caption{\label{fig:3} Left panel: Double slit diffraction pattern with gravitational pull $F=4$. The beam is shot upwards with kinetic energy $E=3$, $m_i= m_g = 1$. The tunneling mark shows the transition to complex quasi time and waves become evanescent, as expected. The curves of constant probability density suffer a bend sideways, producing a fountain-like pattern. Right panel: Same gravitational pull but less kinetic energy $E=2$. Since the energy is variable, we employ here dimensionless variables $2 z \hbar/ F m_i L^4$ (vertical axis) and $x/L$ (horizontal axis).}
\end{figure*}

In fig. \ref{fig:1} we show a comparison of probability densities with and without gravitational pull, obtaining the predicted time deformation due to $F$. The effect can be used to control the distance between diffraction peaks as a function of $\alpha$. In fig. \ref{fig:2} we show a similar comparison between the near zone pattern emerging from a single slit, showing how the focusing point is moved by $\alpha$. The fine details of the pattern near the edges are comparably rich in both cases. In fig. \ref{fig:3} we invert the configuration so as to resemble a fountain of atoms shot upwards; a classical and a tunneling region can be distinguished in these {\it fleur de lis\ }patterns.

Important relations between (\ref{1.14}) and fundamental tests indicated in 1' and 2' can be established. While WEP entails a variation of $\alpha$, UFF only needs a modification of $\beta$. The former case shall be studied in section \ref{sec:5} in terms of small variations. Here, we may see already that $\beta \rightarrow M_i \beta / m_i$ has an effect in the location of the focus, which probes into 1 and 1'. For a more general and fair comparison, we may work out the quasi-time (\ref{1.8}) for $m_i=m_g$, resulting in $ g\tau = \sqrt{v^2 + 2g(z_0-z)}-\sqrt{v^2 + 2g z_0}$, i.e. independent of mass. The accompanying factor in $x_{p,q}$ is $\sqrt{m_i/\hbar}= \sqrt{\beta}$, shown in (\ref{1.11}), which means that this is the only contribution to the modification of the wave under a change of particle species: a rescaling of its argument through the diffusivity, as announced earlier in the Introduction. 
\\

Finally, we should also note that the time-dependent propagation of wave packets due to a linear potential would offer a very simple alternative for treating this problem, since it would only require a three-dimensional initial condition to be evolved along $t$ with a Gaussian propagator. However, since a diffractive screen is on the way, it would not be advisable to simply guess at the solution in the immediacy of the rejecting plates, i.e. an arbitrary choice of $\psi(x,z;t=0)$ around $z=0$. In contrast, we have presented a method that obtains the solution rigorously in full space. There are other diffraction problems stemming from physical situations in which a chopper is part of the dynamics, i.e. a rejecting wall with a slit that has sudden variations in time and space. For the solutions of those problems, it is convenient to follow the method proposed by Bruckner and Zeilinger \cite{zeilinger1997}. Important work related to neutrons was carried out by G\"ahler and Golub in \cite{golub1990, golub1984}. In what follows we would like to show the solution of our stationary problem without these assumptions and dropping the hypothesis of paraxiality and short wavelengths.

\section{ Formal solutions beyond paraxiality \label{sec:4}}

We would like to investigate the solutions of our problem in the region $|z|\ll \hbar/\sqrt{2m_i E}$, $|x| \gg |z|$ and all values of $E$. This corresponds to observation points in the near zone, except for the edges in the case of slits.

In a very general setting, the diffraction kernel can be written in terms known waves. It is important to distinguish the diffraction kernel from the Green's function of the stationary Schr\"odinger equation, since the former does propagate waves along the $z$ axis, while the latter cannot be used for diffraction. We start from  

\bea
\left[-\frac{\hbar^2}{2m_i}\left( \partial_x^2 + \partial_z^2 \right) + V(z) \right] \psi_{\pm}(x,z)= E \psi_{\pm}(x,z).\nonumber \\
\label{2.1}
\eea
As before, the upper wave $\psi_+$ represents a particle falling into the diffractive plate, so (\ref{2.1}) admits separable solutions and we choose $\psi_+(x,z)=\psi_E(z)$ without transversal dependence. This function can be obtained analytically for a number of potentials, so $\psi_E$ is assumed to be a known solution of

\bea
\left[-\frac{\hbar^2}{2m_i} \partial_z^2 + V(z) \right] \psi_{E}(z)= E \psi_{E}(z).
\label{2.1.2}
\eea

Now we build $\psi_-(x,z)$ as a superposition of solutions of (\ref{2.1}) at constant energy $E$; from the separability of the equation, we use the products $e^{ikx} \psi_{\epsilon}(z)$, which satisfy a dispersion relation $E = \epsilon+ \hbar^2 k^2 /2m_i$:

\bea
\psi_-(x,z) =  \int_{-\infty}^{\infty} dk \quad C(k) e^{ikx} \psi_{\epsilon(k)}(z),
\label{2.2}
\eea
\bea
\epsilon(k) \equiv E- \frac{\hbar^2 k^2}{2m_i}.
\label{2.2.1}
\eea
The coefficients $C(k)$ are easily determined by direct evaluation of (\ref{2.2}) at $z=0$, as we now indicate. Since the l.h.s. of (\ref{2.2}) is connected to the wave function coming out of the plate, we write:

\bea
\psi_-(x,0)=\phi_E(x)= \int_{-\infty}^{\infty}dk \, C(k)\, e^{ikx} \psi_{\epsilon(k)}(z=0), \nonumber \\
\label{2.3}
\eea
and $\phi_E(x)$ is the same wave as $\psi_E(z=0)$ blocked in some intervals of $x$ by the structure of the diffraction plate. 
Then, the expression (\ref{2.3}) can be Fourier-inverted to give the coefficient
\bea
C(k)= \frac{1}{2\pi}\int_{-\infty}^{\infty} dx' \, \left(\frac{\phi_{E}(x')}{\psi_{\epsilon(k)}(0)} \right) e^{-ikx'}. 
\label{2.3.1}
\eea
Finally, (\ref{2.3.1}) is replaced in (\ref{2.2}) and the propagated wave is written in terms of a new diffraction kernel
\bea
\psi_-(x,z) = \int_{-\infty}^{\infty} dx' \phi_E(x') K(x-x',z;E)
\label{2.4}
\eea
with 

\bea
K(x-x',z;E) = \int_{-\infty}^{\infty}\frac{dk}{2\pi} \left(\frac{\psi_{\epsilon(k)}(z)}{\psi_{\epsilon(k)}(0)} \right) \, e^{ik(x-x')}.
\label{2.5}
\eea
Our treatment holds for any potential $V(z)$. To fix ideas, let us substitute the Airy functions for the choice $V(z)=Fz$

\bea
K_{\rm Grav}(x-x',z;E) &=& \int_{-\infty}^{\infty}\frac{dk}{2\pi} \left[\frac{ {\rm Ai\ }( \kappa z -\gamma \epsilon(k))}{{\rm Ai\ }( -\gamma \epsilon(k))} \right] \times \nonumber \\ &\times& e^{ik(x-x')},
\label{2.6}
\eea
where $\kappa= (\hbar^{2}F^{5})^{1/3}/2m_i$, and $\gamma=\kappa/F$. In (\ref{2.6}), any normalization factors accompanying the Airy functions cancel out due to the quotient $\psi_{\epsilon}(z)/\psi_{\epsilon}(0)$. This function is itself a solution of the stationary problem with normalization $\psi(0)=1$. 

\subsection{Wave falling in the near zone}

\begin{figure}[t]
\includegraphics[width=9cm]{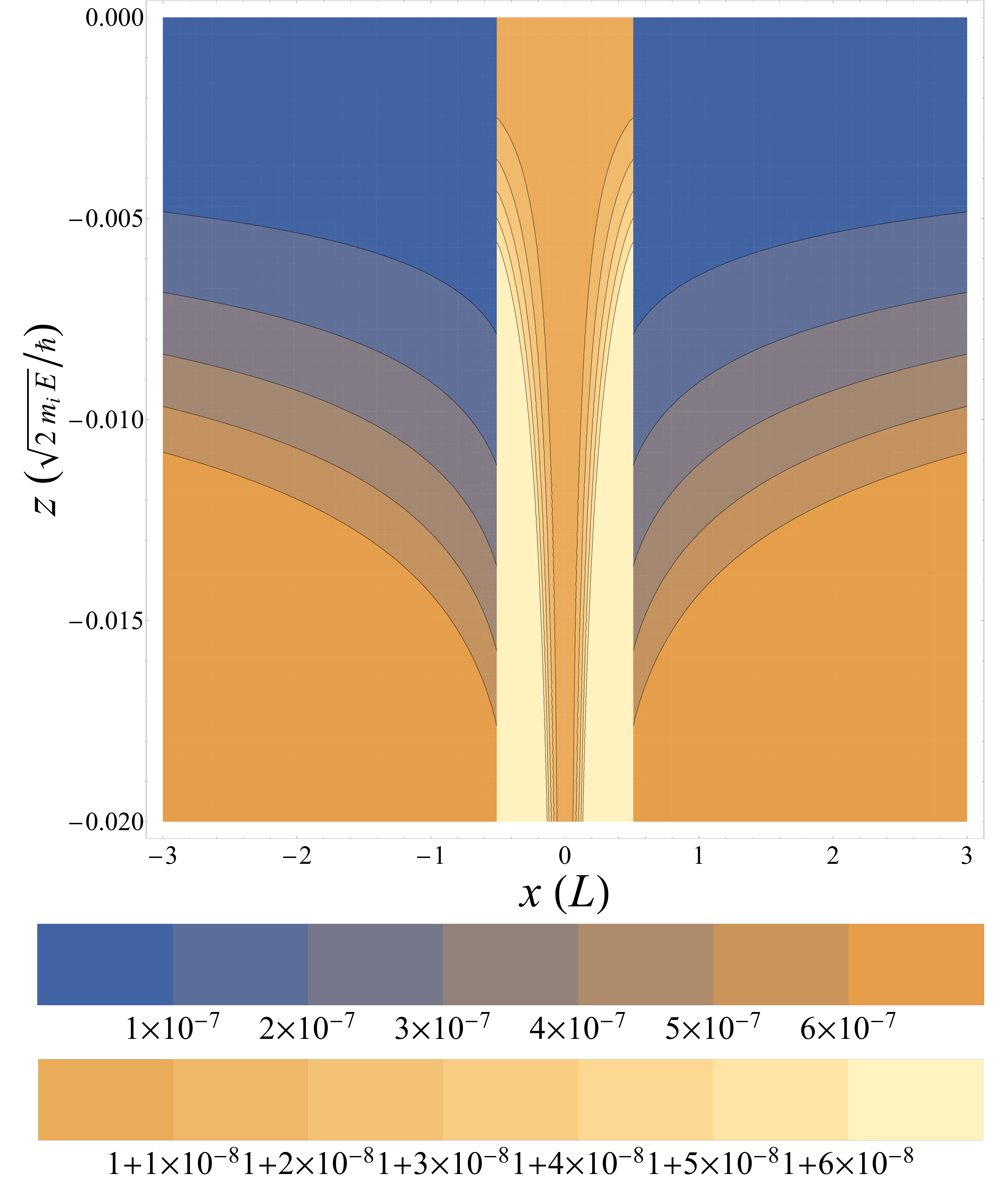} 
\caption{\label{fig:4} A contour plot of $|\psi_-(x,z)|^2$ representing a particle falling from rest, very close to the diffraction plate. As expected, the use of non-paraxial waves produces regular results in the region $z\ll \sqrt{2m_i E}/\hbar$ and $|x\pm L/2|\gg z$. Very small corrections to a Heaviside pulse are present, as shown in the color bar. }
\end{figure}

For a single slit, direct integration of (\ref{2.4}) using (\ref{1.9}) and (\ref{2.6}) obtains

\bea
\psi_-(x,z)&=& \frac{1}{2\pi i \sqrt{L}} \int_{-\infty}^{\infty} \frac{dk}{k} \left[\frac{ {\rm Ai\ }( \kappa z -\gamma \epsilon(k))}{{\rm Ai\ }( -\gamma \epsilon(k))} \right] \times \nonumber \\ &\times &\left[ e^{ik(x+L/2)} - e^{ik(x-L/2)}  \right]. 
\label{2.7}
\eea
Taylor-expanding the Airy quotient for small $z$ must yield good results. The simplest way to proceed is to insert in (\ref{2.7}) the asymptotic form of the Airy function \cite{nist} for large $k$:

\bea
\frac{ {\rm Ai\ }( \kappa z -\gamma \epsilon(k))}{{\rm Ai\ }( -\gamma \epsilon(k))} &\approx & \frac{(-\gamma \epsilon)^{1/4}}{(\kappa z-\gamma \epsilon)^{1/4}} \times \nonumber \\ &\times & \exp \left\{ -\frac{2i}{3} \left[ (\gamma \epsilon-\kappa z)^{3/2}-(\gamma \epsilon)^{3/2} \right] \right\}.\nonumber \\
\label{2.8}
\eea
When $z$ is small, the leading correction comes from

\bea
\frac{ {\rm Ai\ }( \kappa z -\gamma \epsilon(k))}{{\rm Ai\ }( -\gamma \epsilon(k))} \approx 1 + \frac{\kappa z}{4 \gamma \epsilon}.
\label{2.9}
\eea
If the particle (atomic cloud, beam of neutrons, etcetera) drops from the slit at negligible velocity, one has $\epsilon \rightarrow - \hbar^2 k^2 / 2 m_i$ and the wave (\ref{2.7}) reproduces the initial condition plus regular corrections in $z$ and $x$:

\bea
\psi_-(x,z) \approx \phi_E(x) + \frac{\kappa z}{8 \sqrt{L}} \left[ (x-L/2)^2-(x+L/2)^2 \right]. \nonumber \\
\label{2.10}
\eea
where we have used that $\int_{-\infty}^{\infty} dq e^{iq}/q^3 = -i \pi/2$ is the triple antiderivative of the delta distribution evaluated at $1$. This gives various correct limit cases such as $x \rightarrow \infty$ or $0$. As promised, the zone outside of the paraxial region is well captured by this result. We present the contour plots of $|\psi_-(x,z)|^2$ in fig.\ref{fig:4}. No oscillations are observed near the plates.

In order to connect our result with 1 and 1', we stress here that the presence of $\kappa$ in the corrections (\ref{2.10}) contains a mass dependence $m_i^{2/3}$ if $m_i=m_g$, even for vanishing energies. This shows that $|\psi_-(x,z)|^2$ is modified by a change of particle species $m_i \rightarrow M_i$ in the near zone. This also holds for the more general result (\ref{2.7}), whose evaluation in all regions is more challenging.

\begin{table*}[t]
\caption{\label{tab:1}Beam realizations in different energy scales, the resulting focusing points $z_{\rm focus}(0)$ and the corresponding energy- independent sensitivity factors $z_{\rm focus}'(0)$.}
\begin{ruledtabular}
\begin{tabular}{ccccccc}
 & Type & T ( $^{\circ}$K)&$L^*$ (m)
 &$E_{\rm kin}$ (eV) &$z_{\rm focus}(0)$ (m)&$z_{\rm focus}'(0)$ (m)\\
\hline
n & Nuclear & \begin{tabular}{c} --\\20\\300\end{tabular} & $1 \times 10^{-3}$ 
& \begin{tabular}{c} $3.00\times 10^{-7}$\\$2.58\times 10^{-3}$\\$3.87\times 10^{-2}$\end{tabular} & \begin{tabular}{c} $-10.34$  \footnotemark[1]\\$-6.17\times 10^{5}$\\$-2.38\times 10^{3}$\end{tabular} & 3.73 \\ \hline
NH$_3$& Molecular & \begin{tabular}{c} 77\\300\\1200\end{tabular} & $1 \times 10^{-5}$ 
& \begin{tabular}{c} $9.95\times 10^{-3}$\\$3.87\times 10^{-2}$\\$1.54\times 10^{-1}$\end{tabular} & \begin{tabular}{c} $-0.49$\\$-0.98$\\$-1.96$\end{tabular} & $1.07 \times 10^{-5}$  \footnotemark[2] \\ \hline
Cs& Atomic & 1 & $1\times 10^{-4}$ 
& $8.61 \times 10^{-5}$ & $-19.51$ & $6.59$\\
Rb& BEC & $1.7 \times 10^{-7}$ & $1\times 10^{-4}$
& $4.28 \times 10^{-11}$  \footnotemark[3]& $-2.72$ & $2.71$ \\
K& BEC & $5.0 \times 10^{-7}$ & $1\times 10^{-4}$ 
& $4.28 \times 10^{-11}$  \footnotemark[3]  & $-0.57$ & $0.56$ \\

\end{tabular}
\end{ruledtabular}
\footnotetext[1]{Optimal value, experimentally plausible.}
\footnotetext[2]{Decreases sensitivity.}
\footnotetext[3]{Only the free fall energy is considered.}
\end{table*}

\section{Sensitivity estimates in suggested realizations \label{sec:5}}

Here we make a connection with principle 2 and point 2'. In the case of a single slit, the position of the focus can be used to infer the values of $\alpha, \beta$ as functions of the slit width $L$ and the incident kinetic energy $E_{\rm kin}$. The quantity $z_{\rm focus}$ in (\ref{1.14}) also shows a sensible dependence on the variations of fundamental constants that can be used to our favour. The idea then is to probe into possible violations of $m_g = m_i$ for a fixed interferometer, or variations of $g = F/m_g$ for different positions of the apparatus, i.e. a gravitometer.

We start by writing (\ref{1.14}) in terms of its expected value and its possible variation. The constant $\alpha$ suffers a change due to 

\bea
\tilde g = g + \delta g, \quad m_g = m_i + \delta m_g, 
\label{5.1}
\eea
leading to first order corrections of the form
\bea
\alpha = \alpha_0(1+\varepsilon), \quad \alpha_0 = \frac{1}{g}, \quad \varepsilon = -\frac{\delta g}{g} - \frac{\delta m_g}{m_i}.
\label{5.2}
\eea
The constant $\beta=\beta_0$ remains unchanged. When (\ref{5.1}) is substituted in

\bea
z_{\rm focus} = \frac{\alpha E_{\rm kin}}{m_i} - \frac{L^4 \beta^2}{2 \alpha} \left[0.055 + \frac{\alpha}{\beta L^2} \sqrt{\frac{2 E_{\rm kin}}{m_i}} \right]^2
\label{5.3}
\eea
we obtain a first order variation 
\bea
z_{\rm focus} = z_{\rm focus}(0) + \varepsilon \, z_{\rm focus}'(0)
\label{5.3.1}
\eea
with

\bea
z_{\rm focus}(0) = \frac{\alpha_0 E_{\rm kin}}{m_i} - \frac{L^4 \beta_0^2}{2 \alpha_0} \left[0.055 + \frac{\alpha_0}{\beta_0 L^2} \sqrt{\frac{2 E_{\rm kin}}{m_i}} \right]^2 \nonumber \\
\label{5.4}
\eea
and a sensitivity factor
\bea
z_{\rm focus}'(0) = (0.055)^2 \times \frac{ L^4 \beta_0^2}{2 \alpha_0}.
\label{5.5}
\eea
This sensitivity factor is independent of the incident kinetic energy of the beam. It is extremely sensitive to the slit width due to the fourth power of $L$ and it is of a purely quantum-mechanical nature due to $\beta_0$. In the absence of gravity, $g \rightarrow 0$, this factor also disappears. With (\ref{5.4}) and (\ref{5.5}) we may give some predictions depending on various beam realizations as we now discuss.

\subsection{Beam realizations \label{sec:5.1}}

In table \ref{tab:1} we include some numerical values of the focusing point $z_{\rm focus}(0)$ and the factor $z_{\rm focus}'(0)$ for several values of masses and kinetic energies. The goal is to achieve focusing lengths and sensitivity factors in the same order of magnitude and in mesoscopic dimensions. In nuclear physics, thermal neutrons were proposed long ago in connection with quantum mechanical diffraction and the so-called diffraction in time \cite{moshinsky52}. This proposal laid among other successful experiments \cite{cow1, cow2, cow3} which, however, did not involve focusing points. It is interesting to find that Goldemberg and Nussenzveig \cite{nussen} already studied the possibility of observing diffractive effects, reaching the conclusion that not even 20$^{\circ}$K neutrons would allow clean results. Current technology \cite{ultracold1, ultracold2} ensures the existence of 300 neV neutrons that may improve dramatically the resolution of diffraction patterns and with this, our gravitometer. The corresponding wavelength reaches up to 50nm, which is smaller than our proposed aperture value $L^*=$ 1 mm, therefore validating the use of a paraxial approximation in our calculations. To our knowledge, this has not been realized yet and, according to our table, this would be the most promising possibility.

In the molecular domain, a beam of ammonia can be produced at several temperatures, ranging from cryogenic environments \cite{effusion1} at 77$^{\circ}$K to room temperature and even hot effusion (Knudsen) cells at 1200$^{\circ}$K \cite{effusion2}. To compute the incident kinetic energy, we use the $v_{rms}$ value obtained from a Maxwellian distribution leading to the typical $E_{\rm kin}= 3K_{\rm B}T/2$ which, according to historical sources \cite{beam}, induces only 5.5 \% of error. Although the numbers for the focusing point reported in our table are encouraging, the sensitivity factor is very small. There is also a caveat concerning the internal structure of these objects \cite{brezger}; here we have neglected their shape and its influence in transmission peaks through a single slit \cite{shore}. 

A beam of Caesium atoms for atomic clocks represents another interesting possibility. Old standards \cite{cesium} already suggested beam temperatures of few kelvins. Using a conservative figure of 1$^{\circ}$K, our formula yields a focusing distance $\sim 20$ m with a sensitivity factor in a similar order of magnitude, which is acceptable. Previous pedagogical realizations with Potassium atoms and high temperature ovens \cite{potassium} could not spot the near zone effect of focusing.

Finally, we have BECs (Bose-Einstein Condensates), which have been used for gravitometric purposes before. High resolution of the atomic cloud is required \cite{BEC1}. The computation of the incident kinetic energy involves two contributions: the energy acquired in the fall from rest with a time of flight $\sim$ 10 ms and the $v_{\rm rms}$ intrinsic to the cloud, obtained from the Bose-Einstein distribution. Even in this short time of flight, the calculation yields a stronger contribution from the fall, given that a 170 nK Rubidium cloud has a thermal $\lambda = h/\sqrt{2\pi m K_{\rm B}T}$ and therefore a negligible $E_{\rm gas} \sim 1.4 \times 10^{-11}$ eV. A similar result holds for a 500 nK Potassium cloud. 

\subsection{ Non-monochromatic beams and deviations \label{sec:5.2}}

For the most promising case of neutrons, it is important to establish the uncertainty in the location of $z_{\rm focus}(0)$ produced by a beam with a spread in energy. The simplest approach is to consider small uncertainties in $E$, i.e. $\delta E = \delta E_{\rm kin}$, due to various kinetic energies of particles emerging from the furnace. It is advantageous to work with (\ref{5.3.1}) and (\ref{5.4}), because (\ref{5.5}) is independent of the energy. Direct differentiation leads to

\bea
\bigg| \frac{\delta z_{\rm focus}(0)}{\delta E_{\rm kin}} \bigg| = 0.055 \times \frac{L^2 \beta_0}{\sqrt{2 m_i E_{\rm kin}}}
\label{5.2.1}
\eea 
This is valid for any suggested realization in the previous section. We estimate now the width $|\delta z_{\rm focus}(0)|$ for neutrons, using the numbers in table \ref{tab:1}. For $E_{\rm kin} = 3\times 10^{-7}$ eV, the expected uncertainty $\delta E_{\rm kin}$ in energy cannot be worse than $10^{-7}$ eV, regardless of the actual accuracy of reported experiments \cite{ultracold1}. If we use the optimistic $\delta E_{\rm kin} = 10^{-8}$ eV, the following width emerges $|\delta z_{\rm focus}(0)| \sim 0.1$ m. 

The maximum is {\it smeared\ }over a few inches along $z$, so the comparison with $z_{\rm focus}(0)\sim 10$ m reveals a reasonably small region. Detection should involve a grid of counters distributed around the predicted maximum in both directions $\delta x \times \delta z$. Regarding $\delta x$, the main preoccupation should be a proper definition of width that eliminates the ever-growing contributions of wave packet standard deviations (due to the uncertainty principle). This was carefully explained in \cite{case} in connection with the focusing effect of light, including experimental results in \cite{goncalves} as a happy outcome.

\section{Conclusions \label{sec:6}}

We have calculated the deformation effect on diffraction patterns due to external fields, both in semiclassical and fully quantum-mechanical regimes. The relevant effects are of the same nature, namely, that the classical trajectory governs the average of the packet in a new coordinate akin to a classical trajectory. It is worth mentioning that this was done without using Ehrenfest's theorem, circumventing the lack of normalizability of the wave. As a result, the origin of the violation of principle 2, via the factor $F/m_i$, is similar to the classical one, while the sensitivity $z'_{\rm focus}(0)$ is affected by $\hbar$. The other effect, i.e. the violation of 1 produced by $m_i/\hbar$, could be pinned down to a direct correlation between transverse interference effects (such as diffraction in the $x$ coordinate) and the evolution along the fall (quasi-time in $z$). This was done by identifying the field and mass dependences of the global maximum position in the case of a single slit pattern, and a field and mass dependence of the distance between maxima in a ground-based screen in the case of a double slit.

One could argue that this kind of behaviour should also be observable in free falling Gaussian wave packets distributed transversally to the falling motion, since their diffusivity is already mass-dependent. However, keeping track of a Gaussian width along the fall is not as advantageous as having a well defined point of reference. We took advantage of a peculiar focusing effect \cite{moshinsky52, case} produced by slits, in order to spot accurately the position of the global maximum as a benchmark. Then we studied its dependence on $m_i, m_g, E$ and $F$. From this study we were able to suggest experiments in nuclear, atomic and molecular realizations with a twofold purpose: fundamental tests and gravitometry. 

\section{Appendix: Transcendental equation \label{sec:a1}}

We derive here the transcendental equation whose roots are associated with maxima of the single-slit diffraction pattern on axis. We start with equations (\ref{1.7}) and (\ref{1.9})

\bea
\left[- \frac{\hbar^2}{2m_i}\partial_x^2 - i \hbar \partial_{\tau} \right]\tilde \psi(x,z(\tau)) = 0.
\label{a1.7}
\eea

\bea
\psi(x,0)= \tilde \psi(x,z(0)) = \frac{1}{\sqrt{L}} \Theta(L/2 - |x|)
\label{a1.9}
\eea
Eq. (\ref{a1.7}) corresponds to the problem of a free particle with a quasi time $\tau$, therefore we can find the wave function $\tilde \psi (x,z(\tau )))$ by applying the propagator:

\bea
K(x,\tau ;x',\tau ')=\sqrt{ \frac{m_i}{2\pi i\hbar |\tau -\tau '|} }\exp \left[\frac{im_i(x-x')^2}{2\hbar |\tau-\tau '|}\right] \nonumber \\
\label{propagator}
\eea
to the initial condition (\ref{a1.9}), resulting in the following expression:

\bea
\tilde \psi(x,z(\tau))=\sqrt{ \frac{m_i}{2\pi i\hbar |\tau -\tau '|} }\int_{-L/2}^{L/2}dx' \exp \left[\frac{im_i(x-x')^2}{2\hbar |\tau-\tau '|}\right], \nonumber \\
\label{psi propagada}
\eea
this integral can be divided in two parts, with the limits $0<x' <L/2$ and $-L/2 <x'<0$, and we perform the change of variable $t=(x'-x) \sqrt{(m_i)/(2\hbar |\tau-\tau '|)}$ in the former and  $t=(x-x') \sqrt{(m_i)/(2\hbar |\tau-\tau '|)}$ in the latter. Since we are looking for a focusing point along the $z$ axis, we require the condition $x=0$, and so we get:

\bea
\tilde \psi(0,z(\tau)) & = &2\sqrt{ \frac{1}{i\pi}}\fcal(Z), \nonumber \\
Z & = & \sqrt{\frac{m_i}{2\hbar |\tau-\tau '|}} \times \frac{L}{2},
\label{psi fresnel}
\eea
in accordance with (\ref{1.12}). The norm of this function is what we need to maximize, so we solve the equation:
\bea
\frac{d}{d\tau} [\fcal(Z)\fcal^*(Z)]=0. 
\label{por maximizar}
\eea
From the imaginary part of this expression we obtain an identity, but from the real part we finally get our transcendental equation:
\bea
\cos(Z^2)\int_{0}^{Z}dt\,\cos(t^2)+\sin(Z^2)\int_{0}^{Z}dt\,\sin(t^2)=0\nonumber, \\
\label{trasceql}
\eea
and the largest root of this equation gives the numerical result $ \hbar \tau / m_i L^2 = 0.055$, if $\tau'=0$. Further estimates of this constant can be obtained using the asymptotic expansions of Fresnel integrals (Error functions) or the Cornu spiral at an angle of $\pi/4$ in the complex plane, leading to the slightly different results $0.054$ and $0.052$, respectively.

\begin{acknowledgments}
The authors are pleased to thank CONACYT and Sistema Nacional de Investigadores for financial support.
\end{acknowledgments}



\begin{thebibliography}{33}
\bibitem{footnote} This variable corresponds to the classical time of flight at a given height $z$, but since we are dealing with a stationary problem, it does not correspond to time $t$.
\end{thebibliography}

\providecommand{\noopsort}[1]{}\providecommand{\singleletter}[1]{#1}%

\end{document}